\documentclass[12pt,a4paper]{article}

\newcommand{\beq}{\begin{equation}}
\newcommand{\eeq}{\end{equation}}

\begin{document}

\begin{flushright}
hep-th/9903157 
\end{flushright}
\vspace{1.8cm}

\begin{center}
 \textbf{\Large Free Energies and Probe Actions \\ for Near-horizon 
D-branes and D1 + D5 System}
\end{center}
\vspace{1.6cm}
\begin{center}
 Shijong Ryang
\end{center}

\begin{center}
\textit{Department of Physics \\ Kyoto Prefectural University of Medicine
\\ Taishogun, Kyoto 603-8334 Japan}
\par
\texttt{ryang@koto.kpu-m.ac.jp}
\end{center}
\vspace{2.8cm}
\begin{abstract}
By working with the free energy for the type II supergravity 
near-horizon solution of $N$ coincident non-extremal 
D$p$-branes we study the transitions among 
the non-conformal D$p$-brane system, the perturbative 
super Yang-Mills theory and a certain system associated with M theory. 
We derive a relation between this free energy and 
the action of a D$p$-brane probe in the $N$ D$p$-brane background.
Constructing the free energy for the five dimensional
black hole labeled by the D1-brane and D5-brane charges we find the similar
relation between it and the action of a D1 or D5 brane probe in the D1 + D5
brane background. These relations are explained by the massive open strings
stretched between the relevant D-branes.
\end{abstract}
\vspace{3cm}
\begin{flushleft}
March, 1999
\end{flushleft}
\newpage
Recently an important and intriguing duality has been proposed by Maldacena
that states the equivalence between the string/M theory compactified on the
anti-de Sitter space (AdS) and certain superconformal field theory living
on
the boundary of the AdS \cite{JM}. 
This duality was suggested by the fact that
the near-horizon geometries of typical brane configurations 
such as D3-brane, M2-brane and M5-brane are given by the product spaces,
$AdS_5\times S^5, AdS_4\times S^7$ and $AdS_7\times S^4$ respectively. 
For the non-extremal $N$ coincident
D3-branes the thermal gravitational action was constructed to give the free
energy of the four dimensional supersymmetric Yang-Mills (SYM) theory with
$U(N)$ gauge symmety in the limit of 
large $N$ and strong 't Hooft coupling and
further the leading correction in inverse powers of the 't Hooft coupling 
was calculated \cite{GKP,KT,GKT}. The four dimensional SYM free energy was
shown to be related with the supergravity interaction 
potential between the $N$ coincident D3-brane source and a D3-brane probe,
 which is derived by evaluating
the effective DBI action of the D3-brane probe in the near-horizon D3-brane
background \cite{TY}. The supergravity interaction potential is interpreted
as the contribution of the massive states to the SYM free energy in the 
Higgs phase. 

The conformal D3-brane system was extended to the $N$ 
coincident non-conformal D$p$-brane 
system where the D$p$-brane supergravity solution,
that is obtained in the decoupling limit, interpolates between the $p + 1$
dimensional perturbative SYM theory and the different limits of 
string/M theory \cite{IMSY}. For the theories on these different domains
the entropies were calculated and expressed in terms of the energy, the 
volume of the system and so on. They were shown to match near the 
transition regions which are specified 
by the critical energies \cite{BISY}.
Recently combining the Maldacena conjecture with the Matrix theory 
the rich structure of  phase diagram for the 
various near-horizon D-brane systems has been explored and 
described by the entropy and the volume of the system \cite{LMS}.
On the other hand the free energy for the D$p$-brane supergravity solution
was calculated \cite{EW,BR} and the effects of the light spatial winding 
modes and the light temporal winding modes were investigated
\cite{BKR}. Taking account
of these stringy modification and Hagedorn behavior the phase diagram was 
constructed to be described by $g_sN$ and the temperature.

Based on the free energy consideration we will show that the free energy of
the $N$ D$p$-brane system becomes that of the $p + 1$ dimensional 
weakly-coupled SYM theory near the transition region specified by the
critical temperature. Near the strong coupling transition region the 
D$p$-brane supergravity free energy will be shown to match onto the free 
energy of the conformal field theory (CFT) description
or the description associated with M theory. 
From the alternative view point the free energy
for the supergravity description will be reproduced by integrating 
approximately the finite temperature
partition function in the canonical ensemble. We will 
compute the effective DBI action of a D$p$-brane probe in the non-extremal
D$p$-brane background and investigate how it is related with the D$p$-brane
supergravity free energy. 

Further we will consider a system of near-extremal
$Q_1$ D1-branes and $Q_5$ D5-branes in the IIB string theory where the
D5-branes are wrapped on $T^4$. The decoupling limit yields the IIB string
theory on $T^4\times S^3\times AdS_3$ which is dual
 to a certain two dimensional 
CFT \cite{JM}. The obtained six dimensional supergravity solution contains
not only the D1-brane but also a solitonic string which is produced by 
wrapping the D5-brane on $T^4$. Compactifying the string direction leads to
a five dimensional black hole. We will derive the free energy for the 
five dimensional near-horizon black hole
 specified by two charges only and construct
the action of a D1-brane or D5-brane probe in the $Q_1$ D1-brane and 
$Q_5$ D5-brane background. Combining them we will search for a relation
between each thermal probe action and the free energy of the D1 + D5 
system. 

Let us begin to write down the string-frame metric and dilaton for $N$ 
coincident non-extremal D$p$-branes with toroidal 
topology $S^1\times T^p$ for fixing the notation
\begin{eqnarray}
ds^2 &=& \frac{1}{\sqrt{H}}( hd\tau^2 + d\vec{y}^2 ) +\sqrt{H}
\left( \frac{dr^2}{h} + r^2d\Omega_{8-p}^2 \right), \nonumber \\
e^{-2(\phi-\phi_{\infty})} &=& H^{\frac{p-3}{2}},
\label{met}\end{eqnarray}
where $H$ is a harmonic function of the transverse coordinates
\begin{eqnarray}
H &=& 1 + \left( \frac{\tilde{L}}{r} \right)^{7-p}, \nonumber \\
\tilde{L}^{7-p} &=& - \frac{r_0^{7-p}}{2} + 
\left( L^{2(7-p)} + \frac{r_{0}^{2(7-p)}}{4} \right)^{1/2}
\end{eqnarray}
and $h$ is the Schwarzschild-like harmonic function 
$h = 1 - (r_0/r)^{7-p}$. The 'electric' component of the Ramond-Ramond
(RR) $(p + 1)$-form potential is given by
\beq 
C_{012\cdots p} = H^{-1}( 1 - H_0 ),
\eeq 
where 
\beq
H_0 = 1 + \left(\frac{L}{r}\right)^{7-p}, \hspace{1cm}
L^{7-p} = d_p(2\pi)^{p-2}g_sNl_s^{7-p}
\label{hld}\eeq
with $d_p = 2^{7-2p}{\pi}^{(9-3p)/2}\Gamma(\frac{7-p}{2})$ and we use $g_s$
as the asymptotic string coupling, $g_s = e^{\phi_{\infty}}$.
The event horizon at $r = r_0$ yields the Hawking temperature 
\beq
T = \frac{7-p}{4\pi r_0\sqrt{H(r_0)}},
\eeq
which is associated with the period $\beta = 1/T$ of the Euclidean time. 
By compactifying the spatial coordinates $\vec{y}$ on a $p$-dimensional
torus, we have a $D = 10 - p$ dimensional black hole. Its string-frame 
metric is transformed into the Einstein-frame one which gives the ADM mass
\beq
M = \frac{{\omega}_{8-p}V_p}{16\pi G}( (8-p)r_0^{7-p} + 
(7-p)\tilde{L}^{7-p} ),
\label{adm}\eeq
where $G$ is the ten dimensional Newton constant, $G = 8\pi^6g_s^2l_s^8$ 
and ${\omega}_{8-p} = 2\pi^{(9-p)/2} /\Gamma(\frac{9-p}{2})$ is the volume
of the unit $(8-p)$-sphere. The area of the surface at the event horizon 
gives the entropy
\beq
S = \frac{A}{4G_D} = \frac{{\omega}_{8-p}V_p}{4G}r_0^{8-p}\sqrt{H(r_0)}
\label{sh}\eeq
with the $D$ dimensional Newton constant $G_D = G/V_p$, which is also 
expressed as 
\beq
S = \frac{(7-p){\omega}_{8-p}}{16\pi G_D}\beta r_0^{7-p}.
\label{sb}\eeq
Here we take the decoupling limit 
\begin{eqnarray}
{\alpha}' &=& l_s^2 \rightarrow 0, \nonumber \\
\left\{ g_{YM}^2 = (2\pi)^{p-2}g_sl_s^{p-3}, \; U = \frac{r}{{\alpha}'},\;
U_0 = \frac{r_0}{{\alpha}'} \right\} &=& \mbox{fixed},
\label{dcl}\end{eqnarray}
which implies the near-horizon limit for the D$p$-brane solution. In SYM
picture with gauge coupling constant $g_{YM}, U$ is the finite expectation
value of the Higgs. From (\ref{met}) under this field theory limit we have
a blow-up throat manifold $X_{bh}$ with metric
\beq
ds^2 = \left(\frac{r}{L}\right)^{\frac{7-p}{2}}( hd\tau^2 + d\vec{y}^2 )
+ \left(\frac{L}{r}\right)^{\frac{7-p}{2}}\left( \frac{dr^2}{h} + 
r^2d{\Omega}_{8-p}^2 \right),
\eeq
which shows a AdS-like geometry with a varing radius of curvature.
The corresponding mass is obtained from (\ref{adm}) as
\beq
M(X_{bh}) \sim M_{EDp} + \frac{9-p}{2}\frac{{\omega}_{8-p}V_p}{16\pi G}
r_0^{7-p}
\label{mbh}\eeq
with the mass of the extremal D$p$-brane  $M_{EDp} = (7-p){\omega}_{8-p}
V_pL^{7-p}/16{\pi}G$. In Ref. \cite{BKR} the action or free energy of the
near-horizon manifold $X_{bh}$ is presented by
\begin{eqnarray}
I(X_{bh}) &=& \beta F(X_{bh}) = \beta M(X_{bh}) - S(X_{bh}) \nonumber \\
    &=& \beta M_{EDp} - \frac{5-p}{2}\frac{{\omega}_{8-p}V_p}
{16\pi G}\beta r_0^{7-p}
\end{eqnarray}
with
\beq
\beta = \frac{4\pi}{7-p}r_0\left(\frac{L}{r_0}\right)^{\frac{7-p}{2}}.
\label{bt}\eeq
In the limit $r_0 \rightarrow 0$ the manifold $X_{bh}$ becomes a 
blow-up throat of the extremal D$p$-brane solution, which is denoted by 
$X_{vac}$. It has zero entropy and mass $M_{EDp}$ so that 
$I(X_{vac}) = \beta M_{EDp}$. The action difference between $X_{bh}$ and 
$X_{vac}$ yields the free energy for the D$p$-brane supergravity solution
\beq
F = \frac{I(X_{bh}) - I(X_{vac})}{\beta} = - \frac{5-p}{2}
\frac{{\omega}_{8-p}V_p}{16\pi G}r_0^{7-p}.
\label{fr}\eeq

Now we are ready to argue the transition between the supergravity 
description and the perturbative SYM description from the view point of
the free energy. The free energy for the supergravity solution is rewritten
in terms of $g_sN$ and $T$ as
\beq
F = - (5-p)C N^2(g_sN)^{\frac{p-3}{5-p}}(Tl_s)^{\frac{2(7-p)}{5-p}}
V_pl_s^{-p-1},
\label{fc}\eeq
where
\beq
C = \frac{(d_p(2\pi)^{p-2})^{\frac{2}{5-p}}}{2(7-p)(2\pi)^p}
\left(\frac{4\pi}{7-p}\right)^{\frac{2(7-p)}{5-p}}.
\label{cc}\eeq
For $p = 3$ the free energy reads $F = - \frac{{\pi}^2}{8} N^2T^4V_3$
\cite{GKP,KT,GKT}. At a given temperature $T$ which is related with the
non-exremality parameter  $r_0 = U_0\alpha'$ through (\ref{bt}), the 
effective dimensionless coupling constant 
in the corresponding $U(N)$ SYM theory is given by \cite{IMSY,BISY}
\beq
g_{eff}^2 \approx g_{YM}^2NU_0^{p-3},
\label{gef}\eeq
so that it depends on the temperature. Since the supergravity description
can be trusted in the region $g_{eff} > 1$, the transition region is 
characterized by $g_{eff} \sim 1$, that is 
\beq
g_sN \sim (Tl_s)^{3-p}.
\label{gt}\eeq
We would like to see how the free energy behaves at the transition region.
Substituting the relation (\ref{gt}) specifying a critical line in the 
$(g_sN, T)$ plane into the supergravity free energy (\ref{fc}) we have
\beq
F \sim - N^2T^{p+1}V_p.
\label{fy}\eeq
It should be noted that the $l_s$ dependence has been cancelled out.
The obtained expression shows the scaling of the free energy for the 
$p + 1$ dimensional perturbative SYM theory which has $N^2$ degrees of
freedom for the adjoint representation of $U(N)$. 
Since the perturbative SYM theory region $g_{eff} < 1$ is expressed as
\begin{eqnarray}
(g_{YM}^2N)^{\frac{1}{3-p}} < T&,& \hspace{1cm} p < 3 \nonumber \\
T < \frac{1}{(g_{YM}^2N)^{\frac{1}{p-3}}} &,& \hspace{1cm} 3 < p <5
\end{eqnarray}
the perturbative SYM theory gives the high temperature description for
$p < 3$ and the low temperature description for $3 < p < 5$, when 
compared with the supergravity description. Using an
interesting relation $U_0 \approx g_{eff}T$, which is obtained from 
(\ref{gef}) and (\ref{bt}), we make the free energy (\ref{fr}) change into
a simple and suggestive form
\beq
F \sim - (5-p)N^2g_{eff}^{p-3}T^{p+1}V_p.
\label{fge}\eeq
The free energy for the non-conformal D$p$-brane system for $g_{eff} > 1$
is regarded as the strongly-coupled free energy
 for the large $N$ SYM theory with the effective
dimensionless coupling constant $g_{eff}$, which shows the planar
behavior. It becomes the conformal one only when $p = 3.$
As seen above it reduces to (\ref{fy}) obviously near the transition
region $g_{eff} \sim 1$. 

Here we consider the partition function in the canonical ensemble for the
D$p$-brane system. From the elimination of $r_0$ between (\ref{mbh}) and
(\ref{sb}) with (\ref{bt}) the entropy can be expressed in terms of the 
mass as
\beq
S = \gamma \sqrt{N}g_s^{\frac{p-3}{2(7-p)}}V_p^{\frac{5-p}{2(7-p)}}
(M - M_{EDp})^{\frac{9-p}{2(7-p)}},
\eeq
where the positive constant $\gamma$ includes a factor 
$(\frac{1}{9-p})^{(9-p)/2(7-p)}$. Using the energy of D$p$-brane above the
extremality $E = M - M_{EDp}$ we see that the density of states
for the near-horizon D$p$-brane system grows like
$n(E) \sim e^{S(E)}$. The partition function given by 
\beq
Z = \int_0^\infty dEe^{-E/T}n(E)
\eeq
well converges when $(9-p)/2(7-p)<1$, that is $p < 5$. It is interesting
to note that a critical behavior emerges at $p = 5$. In the integration 
the saddle point is estimated with a positive constant $\epsilon$ as
\beq
E_0 = \epsilon (9-p)(\sqrt{N}g_s^{\frac{p-3}{2(7-p)}}
V_p^{\frac{5-p}{2(7-p)}}T)^{\frac{2(7-p)}{5-p}}
\eeq
from which the partition function is approximately evaluated as
\beq
Z \sim \exp \left(\frac{5-p}{9-p} \frac{E_0}{T} \right).
\eeq
We can observe that the free energy $F = -T\log Z$  
reproduces the same expression as (\ref{fc}) including a characteristic
coefficient $(5-p)$. For the peculiar $p = 5$ case the entropy is linearly
rising with energy which is similar to the dynamical string \cite{JMM}.
The free energy shows a logarithmic singular behavior when the temperature
is near the Hagedorn temperature of the effective string,
$T_H \sim 1/\sqrt{g_sN}l_s$.  

Alternatively we use a different parameter 
$\alpha$ accompanied with $r_0$ to give
\beq
\tilde{L}^{7-p} = r_0^{7-p}\sinh^2\alpha, \hspace{1cm}
L^{7-p} = \frac{1}{2}r_0^{7-p}\sinh 2\alpha.
\eeq 
The extremal limit is given by $r_0 \rightarrow 0, \alpha \rightarrow
\infty$ with $L^{7-p}$ fixed. Here we work on the radial coordinate defined
by the shift transformation $\hat{r}^d = r^d + r_0^d\sinh^2\alpha$ with 
$d = 7 - p$, used in Ref. \cite{HS}. The event horizon at $r = r_0$ and the
curvature singularity at $r = 0$ are transformed into an outer event 
horizon specifed by $\hat{r}_+^{7-p} = r_0^{7-p}\cosh^2\alpha$ and an inner
horizon, $\hat{r}_-^{7-p} = r_0^{7-p}\sinh^2\alpha$ respectively.
The RR charge of the $N$ coincident D$p$-branes is
\beq
Q = \frac{\omega_{d+1}}{\sqrt{2}\kappa l_s^{d-4}}d
(\hat{r}_+\hat{r}_-)^{\frac{d}{2}}= 
\frac{\omega_{8-p}}{2\sqrt{2}\kappa l_s^{3-p}}(7-p)r_0^{7-p}\sinh 2\alpha
\eeq
with $\kappa^2 = 8\pi G$, and the extremal limit yields $\hat{r}_+^{7-p},
\hat{r}_-^{7-p} \rightarrow L^{7-p}$. In our notation 
$Q = \omega_{8-p}(7-p)L^{7-p}/\sqrt{2}\kappa l_s^{3-p}$ becomes 
$(2\pi)^{\frac{7}{2}-p}N$. The ADM mass (\ref{adm}) of the starting
non blow-up manifold can be described by
\beq
M = \frac{\omega_{8-p}V_p}{16\pi G}L^{7-p}\left( \frac{2}{\sinh 2\alpha}
+ (7-p)\coth \alpha \right),
\eeq
which is compared to another expression
\beq
M = \frac{D-2}{D-3}ST + Q\Phi
\eeq
with
\beq
\Phi = \frac{V_{10-D}l_s^{2(D-7)}}{\omega_{D-2}(D-3)}
\frac{Q}{\hat{r}_+^{D-3}}.
\eeq
Thus we derive the generalized Smarr formula for the mass of $D$ 
dimensional charged black hole and $\Phi$ can be regarded as the
'electric potential' on the outer event horizon \cite{GM}. 

Furthermore, we will return to analyze the free energy (\ref{fc}) for the
D$p$-brane supergravity solution in both the high and low temperature 
regions. First for the supergravity solution carrying the D2-brane charge
in the high temperature region the 2 + 1 dimensional perturbative SYM 
theory governs the dynamics while the low temperature region is described
by the $N$ coincident M2-brane solution localized in the compact 
dimension in M theory, which gives the CFT with SO(8) R-symmetry 
\cite{NS}. This phase transition takes place at $U_0 \sim
g_{YM}^2$ \cite{IMSY} which gives a transition line $g_s \sim Tl_s 
\sqrt{N}$. From (\ref{fc}) the D2-brane supergravity free energy scales as
$F \sim -N^{5/3}g_s^{-1/3}(Tl_s)^{10/3}V_2l_s^{-3}$. 
By eliminating $g_s$ we see that on the transition line it turns
out to be
\beq
F \sim - N^{\frac{3}{2}}T^3V_2.
\label{fdt}\eeq
In Refs. \cite{KT,GKT} the entropies of the non-dilatonic $N$ coincident
$p$-branes such as D3, M2 and M5 branes were evaluated and each free energy
was shown to be 
\beq
F \sim - N^{\frac{p+1}{2}}T^{p+1}V_p,
\label{fcf}\eeq
which is that of the CFT in $p +1$ dimensions. The expression (\ref{fdt})
is just the form characteristic of the free energy for the 2 + 1 
dimensional CFT. This conformal behavior is also seen directly from 
(\ref{fge}) since the low temperature phase transition is specified by
$g_{eff}^2 \sim N$, which is compared with the high temperature transition
line $g_{eff} \sim 1$. 

For the $N$ coincident D4-branes the supergravity
description is replaced by the SYM theory in the low temperature region.
The high temperature region is 
described by the $N$ coincident M5-branes wrapped on the eleventh 
dimensional circle with radius $R_{11} = g_sl_s$ in M theory. 
The wrapped M5-brane region $g_{YM}^2U_0 > N^{1/3}$ \cite{IMSY}
corresponds to the high temperature region $T > 1/(g_sl_sN^{1/3})$.
The supergravity free energy $F = -CN^3g_sT^6V_4l_s$ 
with $C = 2^73^{-7}\pi^4$ for the D4-branes
in (\ref{fc}) becomes 
\beq
F \sim -N^{\frac{8}{3}}T^5V_4
\label{fdf}\eeq
on the transition line. On the other hand corresponding to
 (\ref{fcf}) with $p = 5$ the 5 + 1 dimensional conformal system on the 
$N$ coincident M5-brane world-volume has a free energy 
$F = -2^63^{-7}\pi^3N^3T^6V_5$ \cite{KT, GKT}.
So the free energy of the wrapped M5-brane is estimated as
$- CN^3T^6V_4g_sl_s$, since its volume is $V_5 = V_4 2\pi R_{11}$. 
From this side the free energy on the transition line becomes (\ref{fdf})
as well. Thus the $N$ coincident D4-branes as well as the $N$ coincident
wrapped M5-branes provide the same free energy expression, which 
covers the whole $g_{eff} > 1$ region. 

The $N$ coincident D1-brane system whose high temperature region is 
described by the two dimensional perturbative 
SYM theory, is also replaced by a two 
dimensional free orbifold CFT \cite{DVV} in the low temperature region.
The orbifold CFT region is characterized by $U_0 < g_{YM}$
\cite{IMSY}, that is, $T < \sqrt{g_s/N} l_s^{-1}$. The free energy
(\ref{fc}) with $p = 1$ is given by $F \sim -N^{3/2}g_s^{-1/2}T^3V_1l_s$
, which becomes
\beq
F \sim -NT^2V_1
\eeq
along the low temperature transition line specified by 
$g_s \sim (Tl_s)^2N$. Its expression just 
shows the scaling of the free energy
for the two dimensional CFT, Eq.(\ref{fcf}) with $p =1$.
This conformal structure also emerges from (\ref{fge}) on the transition
region specified by $g_{eff}^2 \sim N$. Thus in the canonical ensemble we
have demonstrated the matching of the free energies near the transition
line with the temperatures of both sides equally fixed, 
which is compared with the entropy matching with the energies of both 
sides equally fixed that was performed in Ref. \cite{BISY}
from the view point of microcanonical ensemble. 

We further proceed to
the $N$ coincident D0-brane system which in the low temperature region
is replaced by the matrix black hole system in eleven dimensions.
The matrix black hole region is specified by $U_0 < g_{YM}^{2/3}N^{1/9}$
\cite{IMSY}, that is translated into $T < g_s^{1/3}N^{-2/9}l_s^{-1}$.
The free energy for the D0-brane solution scales as
$F \sim -N^{7/5}g_s^{-3/5}(Tl_s)^{14/5}l_s^{-1}$. The substitution of
$g_s \sim (TN^{2/9}l_s)^3$ yields a simple scaling form of the free energy
\beq
F \sim -NT
\eeq
along the critical line. Therefore, the corresponding entropy is just
given by $S \sim N$. In Ref. \cite{BFKS}, the matching of mass-entropy
scaling between the Matrix theory compactified on $T^d$ and the 
Schwarzschild black hole in $(11-d)$ dimensions was exhibited by setting
$S \sim N$. Our result corresponds to the $d = 0$ case. This Matrix theory
behavior arises directly from (\ref{fge}) since the critical line is again
expressed as $g_{eff}^2 \sim N$.

We now turn our attention to  a system that a single D$p$-brane probe
 is separated by a distance $r$ from $N$ coincident non-extremal 
 D$p$-branes (source). The world-volume action of 
a D$p$-brane probe in the D$p$-brane background is given by
\beq
I = T_p\int d^{p+1}\sigma e^{-\phi}\sqrt{\det g_{\mu\nu}\partial X^{\mu}
\partial X^{\nu}} - T_p \int C^{(p+1)},
\label{dbi}\eeq
where $C^{(p+1)}$ represents the RR $(p+1)$-form potential that couples to
the D$p$-brane charge and $\mu, \nu = 0, \cdots, 9$ and $T_p$ is the 
D$p$-brane tension $T_p = 1/(2\pi)^pl_s^{p+1}g_s$. The D$p$-brane probe 
will be considered to be static and parallel 
to the source. Choosing a static gauge and substituting 
the background solution $(\ref{met}) - (\ref{hld})$
into (\ref{dbi}) we have an effective action of a D$p$-brane probe at a
position $r$
\beq
I = T_pV_p \beta H^{-1}( \sqrt{h} - 1 + H_0 - H ),
\eeq
where $V_p$ is the volume of the spatial directions along the D$p$-brane.
In the field theory limit (\ref{dcl}) that also means $L \gg r_0, 
L \gg r$, the free energy $F_{pr} = I/\beta$ of the D$p$-brane probe 
becomes
\beq
F_{pr} = T_p V_p \left(\frac{r}{L}\right)^{7-p}
\left(\sqrt{1-(\frac{r_0}{r})^{7-p}} - 1 \right).
\eeq
The expansion in $r_0/r$ leads to the supergravity interaction potential
between a D$p$-brane probe and the D$p$-brane source. In terms of the two
parameters with energy dimension, $M = (r^{5-p}/L^{7-p})^{1/2}$ and 
$T = (r_0^{5-p}/L^{7-p})^{1/2}(7-p)/4\pi$, which are in a similar form,
the free energy is expressed as 
\beq
F_{pr} = T_pV_p(ML)^{\frac{2(7-p)}{5-p}}
\left(\sqrt{1-(\frac{4\pi}{7-p}\frac{T}{M})^{\frac{2(7-p)}{5-p}}}-1\right).
\label{fpr}\eeq
Here we consider the long distance region $r \gg r_0$ where a D$p$-brane 
probe is far away from the source. Under the corresponding low temperature
expansion in $T/M$ of (\ref{fpr}) the leading term is computed by 
\beq
F_{pr} = -(7-p)CN(g_sN)^{\frac{p-3}{5-p}}(Tl_s)^{\frac{2(7-p)}{5-p}}V_p
l_s^{-p-1}
\eeq
with $C$ defined in (\ref{cc}). Let us denote the free energy for the 
$N$ coincident D$p$-branes (\ref{fc}) by $F(N)$. In the large $N$ limit
we can find a relation 
\beq
F_{pr} \sim F(N+1) - F(N).
\label{sup}\eeq
In the $N+1$ D$p$-brane system one D$p$-brane is far away from $N$ others.
Subtracting the free energy for the $N$ coincident D$p$-branes (source)
from that for the whole $N + 1$ D$p$-brane system we can extract the free
energy for the single D$p$-brane probe in the $N$ D$p$-brane background.
The presence of the D$p$-brane probe at a distance $r$ from the $N$ 
D$p$-brane source gives an expectation value $U = r/\alpha'$ to the Higgs
field and leads to the gauge symmetry breaking 
$U(N+1) \rightarrow U(N) \times U(1)$. The mass of an open string 
stretched between the probe and the source is given by $U/2\pi$. In this 
picture $F(N)$ in (\ref{sup}), the free energy of the $N$ coincident 
D$p$-branes (source) is considered to be given by the massless 
constribution only. Therefore the probe free energy in (\ref{sup})
can be interpreted as the contribution of the massive open strings in the
supergravity side, or alternatively, the contribution of massive SYM modes.
The derivation of (\ref{sup}) on the other hand supports the existence of
the characteristic coefficient $(5-p)$ in the free energy expression 
(\ref{fc}) for the near-horizon D$p$-brane system.

Here we consider a system of $Q_5$ D$5$-branes
and $Q_1$ D$1$-branes in the IIB string theory compactified on $T^4$.
The D1-brane is parallel to the instanton string in six dimensions arising
from the D5-brane wrapped around $T^4$. The low energy dynamics of this 
system is described by the Higgs branch of two dimensional 
$U(Q_1)\times U(Q_5)$ SYM gauge theory. The IR limit of it is given by the
two dimensional (4,4) super CFT with central charge $c = 6Q_1Q_5$
\cite{MD}. We write down the ten dimensional metric for the 
D1 + D5 brane system 
\begin{eqnarray}
ds^2 &=& H_1^{-\frac{1}{2}}H_5^{-\frac{1}{2}}(hd\tau^2 + dy^2) + 
H_1^{\frac{1}{2}}H_5^{-\frac{1}{2}}d\vec{y}^2 +
H_1^{\frac{1}{2}}H_5^{\frac{1}{2}}
\left(\frac{dr^2}{h}+r^2d\Omega_3^2\right), \nonumber \\
H_1 &=& 1 + \left(\frac{r_1}{r}\right)^2, \hspace{1cm} H_5 = 1 + 
\left(\frac{r_5}{r}\right)^2,\hspace{1cm} h = 1 - 
\left(\frac{r_0}{r}\right)^2,
\end{eqnarray}
where $\vec{y}$ are the coordinates on $T^4$ and 
$y$ is the coordinate along the D1-brane, $r_1^2 = 
g_sQ_1l_s^2/v$ and $r_5^2 = g_sQ_5l_s^2$. In the decoupling limit
defined by
\beq
\alpha' \rightarrow 0,\; \hspace{1cm}
\left\{\frac{r}{\alpha'},\; v = \frac{V_4}{(2\pi)^4\alpha'^2},\;
g_6 = \frac{g_s}{\sqrt{v}} \right\} = \mbox{fixed}
\label{dfo}\eeq
with $V_4$ the volume of $T^4$, the D1 + D5 
system reduces to the IIB string theory on $AdS_3\times S^3\times T^4$
with the same radius for $AdS_3$
and $S^3$, which is conjectured to be dual to the two dimensional (4,4)
super CFT. Recently this type of AdS/CFT correspondence has been explored
\cite{SA} to be associated with the three dimesional BTZ black hole and
the $AdS_3$ Chern-Simons gravity. Further compactifying on $S^1$ of length
$2\pi R$ along $y$ direction we get a non-extremal five dimesional 
black hole labeled by two charges, $Q_1$ and $Q_5$.
In the dilute gas condition $r_0 \ll r_1, r_5$, which is contained in the
decoupling limit (\ref{dfo}), the black hole mass and 
entropy are given by
\begin{eqnarray}
M &=& \frac{\pi}{4G_5}\left(r_1^2 + r_5^2 + \frac{r_0^2}{2}\right) = 
\frac{R}{g_s^2l_s^2}\left(g_sQ_1 + vg_sQ_5 + \frac{vr_0^2}{2l_s^2}\right)
,\nonumber \\
S &=& \frac{A}{4G_5} = \frac{2\pi Rr_0}{g_sl_s^2}\sqrt{vQ_1Q_5},
\end{eqnarray}
with the five dimensional Newton constant $G_5 = G/V_42\pi R$ and the
area of the event horizon $A = \pi^2r_1r_5r_0$. In the extremal limit
$r_0 \rightarrow 0$ the entropy vanishes, which is compared to the 
non-zero entropy of the five dimensional extremal black hole with three
charges, $Q_1, Q_5$ and further momentum running along 
the string \cite{HMS,CM}. The non-extremality parameter $r_0$ is linear
to the Hawking temperature as $r_0 = 2\pi r_1r_5T$, that is similarly
seen in the near-horizon D3-brane system. Gathering together we obtain
the free energy 
\beq
F = \frac{R(Q_1 + vQ_5)}{g_sl_s^2} - 2\pi^2Q_1Q_5RT^2.
\eeq
Subtracting the extremal part in the same way as the previous 
D$p$-brane system we have $F(Q_1,Q_5) = -2\pi^2Q_1Q_5RT^2$,
which shows the expected scaling for the free energy of the 1 + 1
dimensional CFT. In deriving $F(Q_1,Q_5)$ we have observed that the
dependences of $g_s$ and $l_s$ are cancelled out. 
It agrees with (\ref{fcf}) with $p = 1, V_1 =2\pi R$ and 
$N = Q_1Q_5$. The combination $Q_1Q_5$ may be associated
with the degrees of freedom in the dual CFT, that is, 
the central charge $c = 6Q_1Q_5$ or the number of the open strings 
connected between D1-branes and D5-branes in a free string gas 
\cite{CM}. So we have an interesting expression 
$F(Q_1,Q_5)/2\pi R = -c\pi T^2/6$. 

The action (\ref{dbi}) with $p = 1$
for a static D1-brane probe in the D1 + D5 brane background can be 
expressed in the decoupling limit as
\beq
I_1 = T_12\pi R\beta H_1^{-1}(\sqrt{h} - 1),
\eeq
where $e^{-2\phi} = H_5/H_1$ is taken into account and $H_1$ is given by
$H_1 \sim g_sQ_1l_s^2/vr^2$. Similar probe actions were studied in 
Ref.\cite{DPS}, where the five dimensional black hole with three charges 
are argued. Under the long distance expansion in $r_0/r$ the leading term
in the free energy of a D1-brane probe at a position $r$ is estimated as
\beq
F_1 = - \frac{Rvr_0^2}{2g_s^2l_s^4Q_1},
\eeq
which is further rewritten in terms of $T$ as 
\beq
F_1 = - 2\pi^2 RQ_5T^2.
\eeq
Comparing the two free energies we can derive a fascinating 
relation
\beq
F_1 = F(Q_1+1, Q_5) - F(Q_1, Q_5),
\eeq
which shows the same subtractive behavior as (\ref{sup}). Thus the free
energy for a D1-brane probe separated from the $Q_1$ D1-brane and
$Q_5$ D5-brane source can be expressed as a difference between the free
energy for $Q_1 + 1$ D1-brane and $Q_5$ D5-brane system and that for
$Q_1$ D1-brane and $Q_5$ D5-brane system. Therefore it can be interpreted
as the contribution of the $Q_5$ massive open strings stretched from the
$Q_5$ D5-branes to the single D1-brane probe. 

Further we repeat a similar analysis for the D5-brane probe in the
same background.
In the action (\ref{dbi}) with $p = 5$ for a D5-brane probe  
the decoupling limit yields
\beq
I_5 = T_5V_5\beta H_5^{-1}(\sqrt{h} - 1)
\eeq
with $V_5 = V_42\pi R$. The leading part of the low temperature 
expansion for the associated free energy is given by
\beq
F_5 = - 2\pi^2 RQ_1 T^2.
\eeq
A similar relation is again encountered as
\beq
F_5 = F(Q_1, Q_5 + 1) - F(Q_1, Q_5),
\eeq
which consistently shows the free energy for the D5-brane probe 
separated from the non-extremal D1 + D5 source. 
The free energy $F_5$ is also characterized by the $Q_1$ massive open
strings stretched between the single D5-brane probe and 
the $Q_1$ D1-branes.

We have analyzed the free energies of the $N$ coincident non-conformal 
D0, D1 and D2 brane systems and observed the free energy matchings by
deriving the perturbative SYM free energies in the high temperature
transition region as well as the free energies of the matrix black hole,
the free orbifold CFT and the CFT with SO(8) R-symmetry in the low 
temperature transition region. We have presented the generalized Smarr
formula for the mass of $D = 10 - p$ dimensional black hole obtained upon
dimensional reduction from the ten dimensional $N$ coincident 
non-extremal D$p$-brane system. The DBI action of the D$p$-brane probe in
the $N$ coincident D$p$-brane background has been shown to be related 
in a subtractive form with the D$p$-brane supergravity free energy and
interpreted as the contribution of massive modes in the SYM picture 
associated with the $U(N+1) \rightarrow U(N)\times U(1)$ 
symmetry breaking. It is interesting to note 
that the characteristic $(5-p)$ factor in the
D$p$-brane supergravity free energy is reproduced by calculating the
finite temperature partition function in 
the canonical ensemble and also plays an important
role in order to obtain the subtractive relation.

The free energy of the five dimensional black hole with two charges 
obtained upon dimensional reduction from the near-horizon D1 + D5
brane system has been constructed and shown to agree with that of 
the two dimensional super CFT. Its free energy is characterized by 
the number of open strings connected between the D1-branes and the
D5-branes. We have also found the subtractive relation between its free
energy and the D1-brane or D5-brane probe action in the D1 + D5 
background. The obtained two relations, which are symmetric each other,
imply a simple and consistent picture that the free energy associated 
with the D1-brane or D5-brane probe action is regarded as 
the constribution of the massive open strings stretched between the 
single probe D1-brane and the source D5-branes or vice versa.

It is worth noting that though we have found the similar subtractive
relations in the two near-horizon systems, the $N$ D$p$-brane system
and the D1 + D5 system, there is a slight difference. The former 
AdS-like geometry is associated with the strongly-coupled $p + 1$ 
dimensional $U(N)$ SYM interacting gauge theory itself, while the
latter AdS itself is associated with the free string gas, or equivalently
the IR limit of the Higgs branch of two dimensional $U(Q_1)\times U(Q_5)$
SYM gauge theory. This difference is reflected in the numbers of 
relevant degrees of freedom and the free energy expressions.
It would be interesting to investigate whether there exist the similar 
subtractive relations between the free energy and the D-brane probe action
in the near-horizon systems for the bound state of the D$p$-branes and 
the D$(p-4)$-branes, and the other bound states in order to gain 
deeper understandings of the duality between the near-horizon system and
the underlying CFT from the thermodynamical view point.

\end{document}